\documentclass[pra,showpacs,twocolumn,floatfix, groupaddress,superscriptaddress,nofootinbib]{revtex4-2}
\usepackage{times,amsmath,amsfonts,amssymb,latexsym}
\usepackage{graphicx,epsf}
\usepackage{subfigure}
\usepackage[shortlabels]{enumitem}
\usepackage{mathtools}
\usepackage{color}
\usepackage{soul}
\newcommand{\be}{\begin{equation}}
\newcommand{\ee}{\end{equation}}
\newcommand{\ba}{\begin{eqnarray}}
\newcommand{\ea}{\end{eqnarray}}
\newcommand\tr{{\mbox{Tr\,}}}
\newcommand{\ignore}[1]{}

\newcommand{\ket}[1]{\left | {#1} \right \rangle }

\newcommand{\bra}[1]{\left \langle {#1} \right | }

\newcommand{\Tr}[0]{\textnormal{Tr}}
\newcommand{\ave}[0]{\textnormal{ave}}
\newcommand{\loc}[0]{\textnormal{loc}}
\DeclareMathOperator*{\Pur}{Pur}

\def\CC{{\rm\kern.24em \vrule width.04em height1.46ex depth-.07ex
    \kern-.30em C}}
\def\P{{\rm I\kern-.25em P}}
\def\RR{{\rm
         \vrule width.04em height1.58ex depth-.0ex
         \kern-.04em R}}

\def\bbbc{{\mathchoice {\setbox0=\hbox{$\displaystyle\rm C$}\hbox{\hbox
to0pt{\kern0.4\wd0\vrule height0.9\ht0\hss}\box0}}
{\setbox0=\hbox{$\textstyle\rm C$}\hbox{\hbox
to0pt{\kern0.4\wd0\vrule height0.9\ht0\hss}\box0}}
{\setbox0=\hbox{$\scriptstyle\rm C$}\hbox{\hbox
to0pt{\kern0.4\wd0\vrule height0.9\ht0\hss}\box0}}
{\setbox0=\hbox{$\scriptscriptstyle\rm C$}\hbox{\hbox
to0pt{\kern0.4\wd0\vrule height0.9\ht0\hss}\box0}}}}
\def\bbbz{{\mathchoice {\hbox{$\sf\textstyle Z\kern-0.4em Z$}}
{\hbox{$\sf\textstyle Z\kern-0.4em Z$}}
{\hbox{$\sf\scriptstyle Z\kern-0.3em Z$}}
{\hbox{$\sf\scriptscriptstyle Z\kern-0.2em Z$}}}}

\usepackage{hyperref}
\hypersetup{colorlinks=true,linkcolor=blue,citecolor=red,filecolor=red,urlcolor=blue,runcolor=blue}

\begin{document}

\title{Localizable quantum coherence}

\author{Alioscia Hamma}
\affiliation{Physics Department,  University of Massachusetts Boston,  02125, USA}

\author{Georgios Styliaris}
\affiliation{Department of Physics and Astronomy, and Center for Quantum Information Science and Technology, University of Southern California, Los Angeles, CA 90089-0484, USA}
\affiliation{Max-Planck-Institut f\"ur Quantenoptik, Hans-Kopfermann-Str. 1, 85748 Garching, Germany}
\affiliation{Munich Center for Quantum Science and Technology, Schellingstraße 4, 80799 M\"unchen, Germany}

\author{Paolo Zanardi}
\affiliation{Department of Physics and Astronomy, and Center for Quantum Information Science and Technology, University of Southern California, Los Angeles, CA 90089-0484, USA}

\begin{abstract}

Coherence is a fundamental notion in quantum mechanics, defined relative to a reference basis. As such, it does not necessarily reveal the locality of interactions nor takes into account the accessible operations in a composite quantum system. In this paper, we put forward a notion of \textit{localizable coherence} as the coherence that can be stored in a particular subsystem, either by measuring or just by disregarding the rest. We examine its spreading, its average properties in the Hilbert space and show that it can be applied to reveal the real-space structure of states of interest in quantum many-body theory, for example, localized or topological states.

\end{abstract}

\maketitle

\section{Introduction}
One of the most striking properties of quantum mechanics is the fact that the state of a quantum system can be expressed as a coherent superposition of different physical states, that is, the eigenstates corresponding to actual measurable values of some observable. Since these eigenstates constitute a basis of perfectly distinguishable states, the coefficients of this linear expansion also depend on the basis. All the purely quantum features are closely related to the presence of quantum coherence, which experimentally manifests itself in  interference and quantum fluctuations~\cite{glauber}. The passage from classical to quantum world is indeed believed to be due to {\em decoherence}~\cite{zurek}. Preserving quantum coherence, and thus fighting decoherence, is one of the most fundamental challenges~\cite{dfs1,dfs2,dfs3} for protocols of quantum information processing~\cite{nielsen}.

The quantitative theory of coherence has witnessed several advances in recent years~\cite{aberg2006quantifying,coh1, coh2} together with its application to the fields of quantum metrology~\cite{qmetro,castellini2019indistinguishability}, quantum foundations~\cite{asymmetry1,asymmetry2}, quantum biology~\cite{qbio} and quantum thermodynamics~\cite{qthermo1,qthermo2}. This approach has also motivated various efforts to extend the quantification of coherence from quantum states to quantum operations~\cite{cohpower1,dana2017resource,korzekwa2018coherifying,theurer2019quantifying,liu2019operational}. In particular, one notion that has surfaced is that of coherence-generating power for a quantum map~\cite{cohpower2,cohpower3,cohpower4,zhang2018coherence}, namely how much coherence can be on average be obtained by a given class of quantum operations.

The notion of coherence per se makes no reference to the {\em locality} of a quantum system~\cite{coh1}. In other words, the basis with respect to which coherence is defined does not necessarily require any underlying tensor product structure of the Hilbert space, as is the case, e.g., for entanglement. On the other hand, every realistic quantum operation is local because of the observables one has access to~\cite{virtual}. To that end, a few approaches towards taking into account the subsystem structure have been proposed~\cite{streltsov2015measuring,ya02015quantum,
radhakrishnan2016distribution,chitambar2016relating,
streltsov2017towards}. One of the basic ideas utilized is to consider incoherent states and operations that, at the same time, respect the underlying local structure of the Hilbert space, obtaining various hybrids between coherence and entanglement.



In this paper, we put forward a notion of {\em localizable coherence}, that is, the coherence that can be stored in a particular subsystem of a quantum system with a given tensor product structure. We investigate different protocols, that involve either disregarding or actively measuring a part of the system, so as to localize quantum coherence in the rest of it. We compute average properties of the introduced quantities in the Hilbert space and investigate the role that measurements, with or without post-selection, have in localizing coherence. Once one has introduced a notion of locality, we use this quantity to characterize the coherence of states that have a particular real space structure, e.g., localized or topological states.

\section{Localizing Coherence}

\subsection{Localizing Coherence by tracing out}\label{traceout}

Consider a (finite dimensional) Hilbert space $\mathcal H = \mathcal H_S \otimes \mathcal H_A$. We see $\mathcal H_S$ as the subsystem in which we want to store coherence, and $\mathcal H_A$ as an environment or an ancillary system. Let  $\dim (\mathcal H) = d = d_S d_A$. Given a quantum state $\rho \in \mathcal B \left( \mathcal H \right)$, a natural way of obtaining a quantum state over $\mathcal H_S $ would be to just trace out the ancillary part and obtain $\rho_S =\Tr_A (\rho)$; then, picking a preferential basis $B_S$ on $\mathcal H_S$, one could simply consider the coherence of the state $\rho_S$ in that basis.

However, it appears immediately that this strategy cannot produce much coherence in  $\mathcal H_S$. The marginal state $\rho_S$ is a state  that has decohered considerably~\cite{zurek} unless $\rho$ is close to separable, which is a rare event~\cite{popescu}. Indeed, with high probability, the marginal state will be typically indistinguishable from the maximally mixed state (for $d_A \gg d_S \gg1$) which is  completely incoherent.

Let us make the above observation more precise. For any measure of coherence $c_B$ with respect to a basis $B$, one can define the coherence of the reduced state $c_{B_S}(\rho_S)$  to represent coherence localized in $S$. We denote
\begin{align}
C_{\Tr,B_S} ^{(S)} (\rho) \coloneqq &  c_{B_S}\left[ \Tr_A (\rho) \right] \,\;.
\end{align}
The connection between coherence and mixedness, as quantified by purity, is illustrated well if one uses in place of the coherence measure $c_B$ the (squared) 2--norm of coherence~\cite{de2016genuine,cohpower2,korzekwa2018coherifying}. The latter is given by
\begin{align}
c_{2,B}(\sigma) \coloneqq  \left\| (\mathcal I -\mathcal D_B) \sigma  \right\|_2^2  = \Pur(\sigma) -\Pur \left[ \mathcal D_B (\sigma) \right] \,,
\end{align}
where $\left\| X \right\|_2 \coloneqq \sqrt{\Tr \left( X^\dagger X \right)}$ denotes the (Schatten) 2--norm, $\mathcal D_{B_S} (X) \coloneqq \sum_k \chi_k X \chi_k$ is the dephasing superoperator, $B_S =\{ \chi_k \}_{k=1}^{d_S}$ denotes a basis on $\mathcal H_S$ consisting of \text{rank-1} orthogonal projectors $\chi_k=\ket{k}\!\bra{k}$, while
$\Pur(\rho) \coloneqq \Tr (\rho^2)$ denotes the purity.\footnote{Notice that the 2-coherence $c_{2,B}$ might fail to satisfy the monotonicity property under the action of the free operations, depending on how one defines the resource theory of coherence (see, e.g.,~\cite{coh2} for more details). Nevertheless, it admits a simple interpretation as an escape probability~\cite{styliaris2019quantum}.} With respect to this measure of coherence, one obtains in terms of purity,
\begin{align} \label{trcoh}
C_{\Tr,B_S} ^{(S)} (\rho) =  \Pur \left(  \Tr_A (\rho) \right)  -\Pur \left( \mathcal D_{B_S}  \Tr_A (\rho) \right) \,\;. 
\end{align}
As it can be seen from the above equation, the purity of the reduced state establishes an upper bound to the coherence of the reduced state.

For a random pure state (i.e., an initial pure state distributed according to the Haar measure $\rho = U \ket{\psi} \bra{\psi} U^\dagger$) the average purity is
\begin{align} \label{avtr}
\overline{\Pur \left[  \Tr_A (U \ket{\psi} \bra{\psi} U^\dagger) \right]}^{\,U} = \frac{d_S + d_A}{d_S d_A + 1}
\end{align}
which implies that, for $d_A \gg d_S$, $\Tr_A (\rho)$ is typically maximally mixed~\cite{physicalstates}. Using this result, a straightforward calculation gives for the coherence
\begin{align} \label{tr_ave_coh}
\overline{C^{(S)}_{\Tr, B_S} \left[  U \ket{\psi} \bra{\psi} U^\dagger) \right]}^{\,U} = \frac{d_S - 1}{d+1} \,\;.
\end{align}
Even for $d_S\simeq d_A$, one obtains an average coherence $\overline{C_{\Tr,B_S} ^{(S)}}\sim 1/d_S$ which is exponentially small in the number of constituents in the $S$ system.

We have seen that the more a state is entangled, the less coherence can be stored in the local system by just tracing out the ancillary part. One can evaluate the relationship between coherence and entanglement by writing a pure state $\rho$ in a Schmidt decomposition. Expressing
\begin{align} \label{schmidt}
\rho =\sum_{a,b=1}^R  c_a c_b^* \ket{\xi_a\eta_a}\!\bra{\xi_b\eta_b} \,\;,
\end{align}
the reduced density matrix reads $\rho_S = \sum_a^R |c_a|^2\ket{\xi_a}\!\bra{\xi_a} $.
The 2--norm of coherence of the reduced state $\rho_S$ is given by
\begin{align}
\label{cohsch2}
c_{2,B_S}(\rho_S) = \sum_a |c_a|^4 - \sum_k \left( \sum_a|c_a|^2 |\langle \xi_a| k\rangle |^2\right)^2 \,.
\end{align}
Recall that two bases are mutually unbiased if the modulus of the inner product between any two basis states is equal to $d^{-1/2}$.
Then, from the above expression, 
it also follows that, for a fixed reduced state $\rho_S$, the coherence $c_{2,B_S}(\rho_S)$ is always maximum over a basis that is unbiased with respect to the Schmidt basis $\{ \ket{\xi_a} \! \bra{\xi_a} \}_a$ (more generally, unbiased to an eigenbasis of $\rho_S$).
Therefore, in order to maximize coherence, one should measure it over a basis that is as unbiased as possible with respect to the Schmidt basis.

One can additionally consider the $l_1$--norm of coherence\footnote{$\| X \|_{l_1} \coloneqq \sum_{ij} |X_{ij}|$ for a matrix $X$.}~\cite{coh1}, which reads
\begin{align}
c_{1,B}(\sigma) \coloneqq \| (\mathcal I - \mathcal D_{B})\sigma\|_{l_1}
\end{align}
($\sigma$ above is understood as a matrix in the $B$ basis) and, for the reduced state, it gives
\begin{align}
\label{cohsch1}
c_{1,B_S} \left( \rho_S \right) =\sum_{k\ne k'} \Big| \sum_{a=1}^R |c_a|^2  \langle\xi_a |k\rangle\langle k'|\xi_a\rangle  \Big| \,\;.
\end{align}
We will see later in section~\ref{sec:toric} that these expressions are useful in the case of quantum states with a particular structure, e.g., topologically ordered states.

\subsection{Localizing coherence by measurement}

Let us now investigate an alternative strategy to localize coherence in $S$ that involves performing an orthogonal measurement on the ancillary system $\mathcal H_A$. After the measurement process, the resulting state is in a product form (some state on $\mathcal H_S$ times an eigenstate of the operator measured on $\mathcal H_A$).
This is a strategy that has been employed to localize entanglement and circumvent the notorious difficulties in measuring entanglement in a mixed state~\cite{le}.  
We  pick some preferred basis $B_A \coloneqq  \{ \omega_i \}_{i=1}^{d_A}$ where the $\omega_i \coloneqq \ket{i} \! \bra{i}$ form a complete set of rank-1 projectors over $ \mathcal H_A$.  A  measurement on $\mathcal H_A$ of a (non-degenerate) observable diagonal in $B_A$ with result ``$i$'' transforms $\rho$ to a product state of the form
\begin{align}
\rho'_i \coloneqq \frac{\Tr_{A} \left( \rho \, I_S\otimes\omega_i \right)}{\Tr\left( \rho \, I_S\otimes\omega_i \right)} \otimes \omega_i \,\;.
\end{align}
This is the result of the measurement where one has retained the information about the outcome $i$. 

For a measurement that is non-selective,
since  each $\rho'_i$ is obtained with probability $p_i = \Tr\left( \rho \, I_S\otimes \omega_i \right)$, the post-measurement state in $\mathcal H$ is
\begin{align}
\rho' =  \mathcal D_{B_A} (\rho) \coloneqq  \sum_i \Tr_{A} \left( \rho \, I_S\otimes\omega_i \right) \otimes \omega_i = \sum_i p_i  \rho'_{i} 
\end{align}
where
\begin{align}
\mathcal D_{B_{A}} (X) = \sum_i I_S \otimes \omega_i X I_S \otimes \omega_i \,\;,\quad  \forall \, X \in \mathcal B(\mathcal H)
\end{align}
is the dephasing superoperator with respect to the basis $B_A$, and similarly $\mathcal D_{B_{S}}$ is the dephasing superoperator in a basis of $\mathcal H_S$. Note that if a basis $B$ of $\mathcal H$ factorizes, i.e., the projectors can take the form $B = B_S \otimes B_A$, then also the (total) dephasing factorizes, namely 
\begin{align}
\mathcal D_{B} = \mathcal D_{B_S}  \mathcal D_{B_A}
\end{align}
and
\begin{align}
\mathcal D_{B_{}} (X) = \sum_{kl} \chi_k\otimes \omega_l X \chi_k \otimes \omega_l \,\;,\,\;  \forall \, X \in \mathcal B(\mathcal H) \,.
\end{align}
In the rest of the paper, we will always assume that the basis factorizes appropriately.

At this point, given a coherence measure $c_B$, we can define the following two quantities: The first one,
\begin{align}\label{locmeas1}
\quad C_{B} ^{(S)} \left( \rho \right) \coloneqq c_{B} \left(    \mathcal D_{B_A} \rho   \right)   \,\;
\end{align}
corresponds to the coherence of the post-measurement state $\rho'$, considered over the whole Hilbert space $\mathcal H$. Notice that the reduced state $\Tr_S(\rho')$ is incoherent.
The second quantity is
\begin{align} \label{average_measure}
C_{\ave,B} ^{(S)} (\rho) \coloneqq \sum_i p_i \,  c_{B_S} \left( \rho'_{S,i} \right) 
\end{align}
where 
\begin{align}
\rho'_{S,i} \coloneqq \Tr_A \left( \rho'_i \right) =\frac{\Tr_{A} \left( \rho \, I_S\otimes\omega_i \right)}{\Tr\left( \rho \, I_S\otimes\omega_i \right)} 
\end{align}
corresponds to the post-selected state in $S$. Therefore the  quantity in Eq.~\eqref{average_measure} corresponds to the average coherence present in each post-measurement state,  restricted to the subsystem $S$.


Using the definitions introduced in Eqs.~\eqref{locmeas1} and \eqref{average_measure}, one could also define the corresponding optimal localizable coherence by taking the supremum over the measurement basis in a given state, or perform the average localizable coherence by Haar averaging over the states, which we will analyze later in section~\ref{avh}.

Let us compare the two protocols $C_{\ave,B} ^{(S)}$ and $C_{B} ^{(S)}$ under some general assumptions for the coherence measure. If the measure $c_B$ is convex it immediately follows that
\begin{align}
C_{B} ^{(S)} \left( \rho \right) \le  \sum_i p_i c_B (\rho'_{S,i} \otimes \omega_i) \,\;.
\end{align}
In addition, if the measure also satisfies  $c_B (\rho \otimes \omega_i) = c_{B_S} (\rho)$ (for all $i$ and states $\rho$), then one immediately gets that
\begin{align}
C_{B} ^{(S)} \left( \rho \right) \le C_{\ave,B} ^{(S)} (\rho) \,\;.
\end{align}
Notice that the measures $c_{1,B}$ and $c_{2,B}$ satisfy both assumptions, hence also the above inequality. 

Let us now compare the above quantities (that involve measurement) with the earlier protocol $C_{\Tr,B_S} ^{(S)} $ of tracing out the ancillary part. For the coherence measure $c_{1,B}$ it holds that
\begin{align}
C_{\Tr,B_S} ^{(S)} \left( \rho \right)	  \le   C_{B} ^{(S)} \left( \rho \right) \,\;.
\end{align}
In fact, the above inequality is true for any coherence measure that is monotonic with respect to  the operation of partial dephasing $\mathcal I \otimes \mathcal D_{B_A}$, and also to partially tracing out part $A$. Indeed, $c_{1,B}$ has both of these properties~\cite{coh1}. Notice, however, that although $c_{2,B}$ also satisfies monotonicity under partial dephasing\footnote{This follows from the fact that the 2-norm is monotonic under unital incoherent operations, such as the partial dephasing considered here.}, it fails to satisfy monotonicity under the partial trace, as it can be checked explicitly by considering a product state.

Notice that the (non-selective) measurement procedure corresponding to $C_{B} ^{(S)}$  will not be able to localize any coherence in the system $S$ if we start with a state that is already incoherent. In fact, if the coherence measure $c_B$ is monotonic with respect to $D_{B_A}$, the resulting coherence $C_{B} ^{(S)} \left( \rho \right)$ is upper bounded by $c_B(\rho)$.

We now regard the question of finding the basis $B_S$ that maximizes each of the localizable coherence by measurements $C_{B} ^{(S)}(\rho)$ and $C_{\ave,B} ^{(S)} (\rho)$, for fixed $B_A$ and $\rho$. The optimal basis turns out to be simple for the case when $\{ \rho'_{S,i} \}_i$ are mutually commuting and the coherence measure is $c_{2,B}$. Then, as we show in Appendix~\ref{appendix}, both localizable coherences become maximal for any $B_S$ that is unbiased with respect to $B'_S$ which simultaneously diagonalizes $\{ \rho'_{S,i} \}_i$.
However, we expect the answer to be more complicated for general scenarios.

Let us now  invoke the above result to make a connection with entanglement. As a first simple example, let us consider a separable pure state $\ket{\psi} = \ket{\xi} \ket{\eta}$. For any choice of the measurement basis $B_A$, the assumption of mutually commuting $\{ \rho'_{S,i} \}_i$ is trivially satisfied, and hence an optimal $B_S$ is given by any basis that is unbiased to the single element $\ket{\xi} \! \bra{\xi}$.

One can also consider as an example the opposite limit of a maximally entangled pure state, i.e., as in Eq.~\eqref{schmidt} with $d_A = d_B = \sqrt{d}$ and $c_a = d^{-1/4}$. For a measurement basis $B_A$ related with the Schmidt basis $\{ \ket{\eta_a}\!\bra{\eta_a} \}_a$ of the ancillary system by a quantum Fourier transform $\mathcal F$, it follows that the optimal basis on the system part is given by the Schmidt basis itself $B_S = \{ \ket{\xi_a}\!\bra{\xi_a} \}_a$.
This is because all $\{ \rho'_{S,i} \}_i$ are mutually commuting and, in fact, diagonal in the basis $\mathcal F (B_S)$.\footnote{One way to see this is by expressing $\rho'_{S,i}$ in the $\{ \ket{\xi_a}\!\bra{\xi_a} \}_a$ basis; the resulting matrix is circulant (for all $i$) and hence diagonalizable by a Fourier transform.} Since $B_S$ and $\mathcal F (B_S)$ are unbiased, the claim follows.

\section{Spreading of localizable coherence}

Consider a local quantum system $\mathcal H_\Lambda=\otimes_{x\in\Lambda}\mathcal H_x$ on a lattice $\Lambda$ endowed with graph distance $d(x,y)$ and with each local system a $d$--level system $\mathcal H_x\simeq \mathbb C^d$. We will assume that the dynamics is described by a local Hamiltonian, that is, a Hamiltonian sum of local operators $H=\sum_X\Phi_X$ where $X\subset\Lambda$ and the operators $\Phi_X$ are bounded hermitian operators on $\mathcal H_X = \otimes_{x\in X}\mathcal H_x$. The map $\Phi:X\mapsto \Phi_X$ is the {\em interaction map} that specifies the physical interactions between the particles in the system (including one-body terms). The locality of the subset of sites $X\in\Lambda$ is specified by a bound on the number of sites in $X$, that is, $|X|< R$ and the maximum distance between two sites in $X$, that is, $\mbox{diam}(X) =\max_{x,y\in X} d(x,y)<r$. The number $R$ represents a bound to the maximum number of bodies in an interaction, while $r$ specifies the maximum distance at which bodies can interact.   In this model, correlations spread out with a maximum speed given by the Lieb-Robinson bounds~\cite{LR, Hastings,hastings2006spectral}. In this section we investigate whether also localizable coherence spreads with a given speed. 

In order to establish a connection with our previous setup, we consider a tripartition of the Hilbert space $\mathcal H= \mathcal H_A \otimes  \mathcal H_C \otimes \mathcal H_S$. Here, $\mathcal H_S$ denotes the Hilbert space of the system in which we want to localize coherence. Let the regions $A$, $S$ be separated by a distance $l$. The localizable coherence in $S$ at the time $t$ depends on the details of the initial state $\rho_0$ and on the dynamics, dictated by the Hamiltonian. Unitary evolution will bring the state from $\rho_0$ to $\rho_t = U\rho U^\dagger$ and we would like to investigate in this state the localizable coherence at $S$. What happens if someone at $A$ performs a local quantum operation on the initial state $\rho$? In what follows, we focus for concreteness on the localizable coherence associated with $c_{2,B}$.

The Lieb-Robinson bounds imply that it is impossible to send signals (up to an exponential tail) from $A$ to $S$ outside the light cone. Here we show that also $C_{\Tr,B_S} ^{(S)} (\rho) $ spreads ballistically according to the maximum speed of signaling. On the other hand, a similar result fails to hold in general for both $C_{B} ^{(S)}$ and $C_{\ave,B} ^{(S)}$ that are associated with measurements.


We now make the above claims precise. Let the initial state of the total system be $\rho_0$ and assume some quantum operation is performed on $A$. Then we try to localize coherence on  $S$ after some time $t$. The quantum operation $\mathcal T_A$ will be described by a CPTP map with support on $A$, i.e., its Kraus operators are of the form $\tilde M_A^i  \coloneqq M_A^i \otimes I_{\bar A}$ $\forall i$, where $\bar X$ thereafter denotes the complement of a region $X$. We can therefore define the input state as in that case as
\begin{align}
\rho'_{0} = \mathcal T_A (\rho_0)
\label{rho_k}
\end{align}
Finally, let  $U$ denote the unitary evolution operator to the time $t$ generated by our local Hamiltonian and also $\rho_t$, $ \rho'_t$ the corresponding time evolved states.

Our first result is that a Lieb-Robinson type bound holds for the localizable coherence $C_{\Tr,B_S} ^{(S)} $, namely that 
\begin{align} \label{LRlocalizable}
\left| C_{\Tr,B_S} ^{(S)}  \left( \rho_t \right) - C_{\Tr,B_S} ^{(S)}  \left(  \rho'_t \right) \right| \le  c \exp \left( - \mu l \right) \left[  \exp \left( s \left|  t \right| \right) - 1 \right]   \,\;,
\end{align}
where $c$, $\mu$ and $s$ are positive constants. In particular, for the case of $\rho_0  =  I /d$, the above inequality reduces to
\begin{align}
 C_{\Tr,B_S} ^{(S)}  \left(  \rho'_t \right)  \le  c \exp \left( - \mu l \right) \left[  \exp \left( s \left|  t \right| \right) - 1 \right]   \,\;,
\end{align} 
expressing the fact that a state that is maximally mixed everywhere except possibly at the region $A$ will have exponentially small localizable coherence $C_{\Tr,B_S} ^{(S)} $ outside the light cone.

Let us derive Eq.~\eqref{LRlocalizable}. We begin by first noticing that the function $c_{2,B}(\rho)$ is Lipschitz continuous, namely for any two states it holds that
\begin{align} \label{lipschitz}
\left| c_{2,B}(\rho_1) -  c_{2,B}(\rho_2) \right| \le 2 \left\| \rho_1 -\rho_2  \right\|_2 \,\;.
\end{align}
This follows from the sequence of inequalities (we set $\mathcal Q_B \coloneqq \mathcal I - \mathcal D_B$),
\begin{align*}
&\quad  \,\left| c_{2,B}(\rho_1) -  c_{2,B}(\rho_2) \right| = \left|   \left\|  \mathcal Q_B (\rho_1) \right\|_2^2 -  \left\|  \mathcal Q_B (\rho_2) \right\|_2^2 \right| \\
&= \left(   \left\|  \mathcal Q_B (\rho_1) \right\|_2 +  \left\| \mathcal  Q_B (\rho_2) \right\|_2 \right) \left|   \left\|  \mathcal Q_B (\rho_1) \right\|_2 -  \left\| \mathcal  Q_B (\rho_2) \right\|_2 \right| \\
&\le 2 \left|  \left\| \mathcal  Q_B (\rho_1) \right\|_2 -  \left\| \mathcal  Q_B (\rho_2) \right\|_2 \right| \\
&\le 2 \left\| \mathcal  Q_B (\rho_1) - \mathcal Q_B (\rho_2) \right\|_2  \le 2 \left\|  \rho_1 - \rho_2 \right\|_2  \,\;.
\end{align*}

To show Eq.~\eqref{LRlocalizable}, we need to show that $\left\|  \Tr_{\bar S} \left( \rho_t -  \rho'_t \right) \right\|_2$ is exponentially small outside the light cone. Since
\begin{align*}
\left\|    \rho_1 - \rho_2  \right\|_2 \le \left\|   \rho_1 - \rho_2 \right\|_1 = \sup_{\left\| O  \right\|_\infty = 1 } \Tr \left[ O \left(  \rho_1 - \rho_2  \right)  \right]
\end{align*}
let us consider $  \Tr_S \left[ O_S \Tr_{\bar S} \left(  \rho_t - \tilde \rho_t  \right)  \right]$. We have, 
\begin{align*}
& \Tr_S \left[ O_S \Tr_{\bar S} \left(  \rho_t - \tilde \rho_t  \right)  \right] 
= \Tr \left[ \tilde O_S U \left(  \rho_0- \tilde \rho_0 \right)  U^\dagger \right] \\
= & \, \Tr \left[ \tilde O_S (t) \left(  \rho_0 - \tilde \rho_0  \right) \right]   =  \Tr \left ( (\mathcal I - \mathcal T^*_A) [ \tilde O_S (t) ] \rho_0  \right) \\
 \le & \, \big\| (\mathcal I - \mathcal T^*_A)  \tilde O_S (t)   \big\|  _{\infty}  =   \big\| { \textstyle \sum_i } \tilde M_{A}^{i \, \dagger}  \big[  \tilde M_{A}^{i}  , \tilde O_S(t) \big]   \big\|_{\infty} \\
 \le & \, { \textstyle \sum_i  }  \big\| \big[  \tilde M_{A}^{i}  , \tilde O_S(t) \big]   \big\|_{\infty}  \,\;,
\end{align*}
where above we denote $\tilde O_S (t) \coloneqq U^\dagger O_S \otimes I_{\bar S} U$, while in the last step we have used the fact that the trace preserving condition for $\mathcal T_A$ implies that $\big\| \tilde M_{A}^{i}  \big\|_\infty \le 1$ $\forall i$. We therefore have, by combining the above inequality with Eq.~\eqref{lipschitz}, that
\begin{align*}
\left| C_{\Tr,B_S} ^{(S)}  \left( \rho_t \right) - C_{\Tr,B_S} ^{(S)}  \left(  \rho'_t \right) \right| \le  2 \sup_{\left\| O  \right\|_\infty = 1 }{ \textstyle \sum_i  }  \big\| \big[  \tilde M_{A}^{i}  , \tilde O_S(t) \big]   \big\|_{\infty} \,.
\end{align*}

Each of the above commutators satisfies a Lieb-Robinson bound of the form
\begin{align}
\big\| \big[  \tilde M_{A}^{i}  , \tilde O_S(t) \big]   \big\|_{\infty} \le 2 \left\|   O_S \right\|_\infty \left| S \right| \exp \left( - \mu l \right) \left[ \exp \left( s \left|  t \right| \right) - 1 \right] 
\end{align}
hence we obtain Eq.~\eqref{LRlocalizable} for $c = 4 d_A^2 \left| S \right|$, where the positive constants $s$ and $\mu$ (specifying the Lieb-Robinson velocity) depend on the details of the Hamiltonian and the lattice.

As mentioned earlier, the localizable coherence by measurement (selective or not) does not admit a similar Lieb-Robinson type bound. Indeed, considering an initial state that is separable $\rho_0 = \rho_0^{S} \otimes \rho_0^{AC}$, it is easy to see that the difference $\left| C_{B} ^{(S)}  \left( \rho_0 \right) - C_{B} ^{(S)}  \left(  \rho'_0 \right) \right| \ne 0$, hence no bound analogue to Eq.~\eqref{LRlocalizable} exists for this quantity, and similarly for $C_{\ave,B} ^{(S)}$ .

\section{Average localizable coherence}\label{avh}

As we have seen in section~\ref{traceout}, if we obtain a reduced state to the system $S$ by tracing out the ancillary part $A$, it is expected that this reduced state will not have much coherence in the large Hilbert space dimension limit; typically states are maximally entangled~\cite{popescu}. In this section, we investigate the average value of localizable coherence in the Hilbert space by means of measurement, using the definitions Eqs.~(\ref{locmeas1},\ref{average_measure}). These results will prove useful to understand the local coherence structure of interesting quantum many-body states, such as many-body localized (MBL)~\cite{basko_metal-insulator_2006,pal_many-body_2010,nandkishore_many-body_2015} states or topologically ordered states.

\subsection{Average over global pure states}

In this section, we compute the average of the localizable coherences $C_B ^{(S)},  C_{\ave,B} ^{(S)}$ over the pure states $\rho$ in the Hilbert space according to the Haar measure. 
Since we are interested in average properties, we will be again using the $l_2$--norm measure of coherence $c_{2,B}$. In the following we will always assume that the coherence basis $B = B_S \otimes B_A $ factorizes. The two measures of coherence \eqref{locmeas1} and \eqref{average_measure} then read
\begin{subequations}
\begin{gather}
C_B ^{(S)} (\rho)  = \left\| \left(   \mathcal D_{B_A} - \mathcal D_{B}  \right) \rho   \right\|_2^2     = \Pur \left( \mathcal D_{B_A} \rho \right) - \Pur \left( \mathcal D_{B} \rho \right) \\
C_{\ave,B} ^{(S)} (\rho) = \sum_i p_i  \left[ \Pur\left( \rho'_{S,i} \right) - \Pur\left( \mathcal D_{B_S}\rho'_{S,i} \right)   \right]  \,\;.
\end{gather}
\end{subequations}
After a short calculation one can also obtain the alternative forms
\begin{subequations}
\begin{align} \label{main_measure}
C_B ^{(S)} (\rho) 
&=   c_{B_S} \left( \sum_i p_i \rho'_{i} \right)
= \sum_i p_i^2 c_{B_S} (\rho'_{S,i})\\
&= \sum_i p_i^2 \left(  \Pur(\rho'_{S,i}) - \Pur(\mathcal D_{B_S} \rho'_{S,i}) \right) \,\;.
\end{align}
\end{subequations}

The average over the states $\rho$ can be performed in many different scenarios. To start, we can average uniformly over all the pure states in the Hilbert space. To this end, we write $\rho$ as $\rho_U = U \ket{\psi_0}\bra{\psi_0}U^\dagger$ for a generic reference state $\ket{\psi_0}$. The average over $\rho$ then becomes the Haar average over the unitary group~\cite{physicalstates}.

We start with calculating $\overline{C_B ^{(S)} (\rho)}^\rho$.  To this end, we double the Hilbert space as
\begin{align}
\mathcal H _S \otimes \mathcal H_A \mapsto \mathcal H_{S} \otimes \mathcal H_{A} \otimes \mathcal H_{S'} \otimes \mathcal H_{A'} = \mathcal H \otimes \mathcal H' \,.
\end{align}
We will denote as $S$ the swap operator between $\mathcal H$ and $\mathcal H'$,  acting  on the corresponding basis vectors of $\mathcal H^{\otimes 2}$ as $S\ket{ij}=\ket{ji}$, and also denote $S_{X}$ the swap operator between the $X,X'$ partitions of the doubled system. Recall also that the total swap $S$ factorizes as $S=S_S\otimes S_A$. We can then write the useful identities
\begin{align}
\Tr(X^2) = \Tr(S X \otimes X)  \,\; \quad  \forall \, X \in \mathcal B(\mathcal H) 
\end{align}
and
\begin{align}
\overline{\rho_U ^{\otimes 2} }^{U}  = \frac{1}{d(d+1)} \left( I + S \right)  \,\;.
\end{align}
Exploiting the above identities we obtain
\begin{align}
\nonumber
 \overline{C_B^{(S)}\left(  \rho_U \right)}^{\,U}  
& = \overline{ \Tr \left[ S_{S} \otimes S_{A} ( \mathcal D_{B_{A}}^{\otimes 2}- \mathcal D_{B}^{\otimes 2} ) U^{\otimes 2} \psi_0^{\otimes 2} (U^\dagger) ^{\otimes 2}  \right]}^{U}  \nonumber \\
 & =  \Tr \left[  S_{S} \otimes S_{A} ( \mathcal D_{B_{A}}^{\otimes 2}- \mathcal D_{B}^{\otimes 2} )\frac{I  + S_{S} \otimes S_{A}}{d(d+1)}  \right] \nonumber\\
 & = \frac{1}{d(d+1)} \Tr \left[S_{S} \otimes S_{A} \left( S_{S  } \otimes P_{A}  - P_{S} \otimes P_{A} \right)  \right] \nonumber \\
 & = \frac{1}{d(d+1)} \Tr \left[I_{S}^{\otimes 2}\otimes P_{A} - P_{S} \otimes P_{A} \right] \nonumber \\
 & = \frac{(d_S)^2 d_A - d_S d_A}{d(d+1)} = \frac{d_S - 1}{d+1} 
\end{align}
where we used the notation $\psi_0 \coloneqq \ket{\psi_0} \! \bra{\psi_0}$, also $P_{A} \coloneqq \sum_{i=1}^{d_A} (\ket{i} \bra{i})^{\otimes 2} \in \mathcal B( \mathcal H_A \otimes \mathcal H_{A'})$ and similarly for $P_{S}$. In the third equality we used the fact that $( \mathcal D_{B_{A}}^{\otimes 2}- \mathcal D_{B}^{\otimes 2} ) I =0$ together with $\mathcal D_{B_{A}}^{\otimes 2} (S_{S} \otimes S_{A}) = S_{S  } \otimes P_{A} $  and $\mathcal D_{B_{}}^{\otimes 2} (S_{S} \otimes S_{A}) =P_{S  } \otimes P_{A} $.  In the fourth equality we used  $S_{A,S}^2= I^{\otimes 2}_{A,S}$ and $S_{A,S}P_{A,S}=P_{A,S}$.

In the limit of large Hilbert space dimension $d$, we have  $\overline{C_B^{(S)}}\simeq 1/d_A$. We see that this scheme of measurement returns on average exactly the same coherence as in the case of tracing out, see Eq.~\eqref{tr_ave_coh}.

In the limit $d_S \to d$ (and hence $d_A \to 1)$ we recover the result from~\cite{cohpower2} about average coherence of Haar distributed pure states, that is,
\begin{align} \label{average_coherence}
\overline{c_B (\rho_U)}^{\,U} = \frac{d-1}{d+1} \,\;.
\end{align}

Now we calculate the global Haar average for  $\overline{C_{\ave,B}^{(S)}\left(  U {\psi_0}  U^\dagger \right)}^{\,U}$. In order to perform this calculation, it is convenient to write
\begin{align*}
C_{\ave,B} ^{(S)} (\rho) =  
\sum_i \frac{1}{p_i}  \Tr \left( S_{S} ( \mathcal I^{\otimes 2} -  \mathcal D_{B_S}^{\otimes 2}) [\rho (I_S\otimes\omega_i)]^{\otimes 2}  \right)  .
\end{align*}
This calculation is  more challenging because of the presence of the probability factor $p_i^{-1}$ in the above equation. We now argue can substitute to this value its mean, with an error that becomes irrelevant for large Hilbert space dimension. On average, the probability factor for a given result ``$i$'' takes the value
\begin{align}
 \overline{p_i(U)}^{\,U} = \overline{\Tr\left( \rho_U I_S\otimes \omega_i \right)}^{\,U} = 1/d_A 
\end{align}
This average value is also typical. Indeed, we can invoke Levy's lemma~\cite{watrous2018theory} to bound the probability of having a result different from the average. The function $p_i =\tr (\rho I_S\otimes \omega_i )$ is a function from the $(2n-1)$-dimensional sphere $S^{2n-1}$ to the interval of real values $[0,1]$. Moreover, this function is Lipschitz continuous with Lipschitz constant $\eta =1$ since the maximum difference in probabilities is bounded by one. We can then apply L\'evy lemma with error $\epsilon =d^{-1/3}$ and obtain
\begin{align}
\text{Pr}\left( |p_i(\rho)-\frac{1}{d_A} |\ge d^{-1/3} \right)\le 3\exp\left(- \frac{d^{1/3}}{25\pi}\right)
\end{align}
which shows measure concentration of the function $p_i$. 
At this point, we are justified to use a ``mean field'' approximation in the $\overline{C_{\ave,B}^{(S)}( \rho_U )}^{\,U}  $ calculation by substituting $p_i \approx 1/d_A$, which we expect to be accurate for $d \gg 1$.

Computing the average we obtain
\begin{align}
& \overline{C_{\ave,B}^{(S)}(\rho_U)}^{\,U}  = \nonumber \\ & = \overline { \sum_i \frac{1}{p_i(U)}  \Tr \left( S_{S} (\mathcal I - \mathcal D_{B_S}^{\otimes 2}) ( \rho_U I_S\otimes  \omega_i)^{\otimes 2}  \right) } ^{\,U} \nonumber \\
\nonumber& \approx  d_A   { \sum_i \Tr \left( S_{S} (\mathcal I - \mathcal D_{B_S}^{\otimes 2})  \overline{\rho_U ^{\otimes 2}}^U(I_S\otimes  \omega_i)^{\otimes 2}  \right) } \nonumber\\
 & = \frac{d_A}{d(d+1)} \Tr \left( I_{S}^{\otimes 2}\otimes P_{A} - P_{S}\otimes P_{A} \right) 
\end{align}
and thus 
\begin{align}
\overline{C_{\ave,B}^{(S)}(\rho_U)}^{\,U}  \approx  \frac{ d_S-1}{d_S + 1 / d_A} \,\;.
\end{align}
Notice that, in view of the mean field approximation, this result is just $d_A \overline{C_B^{(S)}\left(  \rho_U \right)}^{\,U} $. 
We can see that in the large $d$ limit we obtain a coherence of order one (e.g., in the limit of $d_A\to \infty$). Moreover, if  $d_A=1$, we then recover Eq.~\eqref{average_coherence}.

\subsection{Average over factorized states}

In this section, we 
consider an initial product state $\ket{\psi_0} = \ket{\psi_0}_S\otimes\ket{\psi_0}_A$ separable in the $(S,A)$ bipartition. We are interested in computing the average localizable coherence to $S$ obtainable by measurement without post-selection, namely,  $\overline{C_B ^{(S)} (\rho)}^\rho$. The density matrix  $\psi_0 =\ket{\psi_0}\bra{\psi_0}$  is of course of the form $\psi_0 = \psi_{0,S}\otimes\psi_{0,A}$. In the following we want to average over all the separable states in this partition according to the Haar measure. To this end, we write $\psi_{0,U} = U_S\psi_{0,S}U_S^\dagger\otimes U_A\psi_{0,A}U_A^\dagger$ with $U=U_S\otimes U_A$ and the Haar average is performed over the unitaries of the form $U_S\otimes U_A$.

The calculation will proceed similarly as before. Performing the average 
\begin{multline}
 \overline{(U_S\otimes U_A)^{\otimes 2} (\psi_{0,S}\otimes\psi_{0,A})^{\otimes 2} (U_S^\dagger\otimes U_A^\dagger) ^{\otimes 2}  }^{U_S\otimes U_A}  =  \\
 = \frac{(I_{S}^{\otimes 2} + S_{S}) \otimes (I_{A}^{\otimes 2} + S_{A})}{d_S(d_S+1) d_A (d_A + 1)} \,\;,
\end{multline}
we obtain 
\begin{align}
\nonumber
\overline{C_B^{(S)} \!\left(  \rho_U \right)}^{\,U}  \!\!\!\!
&= \frac{ \Tr \big( S ( \mathcal D_{B_A }^{\otimes 2} \!\!- \!\mathcal D_{B}^{\otimes 2}) [(I_{S}^{\otimes 2} + S_{S}) \otimes (I_{A}^{\otimes 2} + S_{A}) ] \big)}{d_S(d_S+1) d_A (d_A + 1)}\\
\nonumber & = 2 \frac{\Tr [I_{S}^{\otimes 2}\otimes P_{A} - P_{S} \otimes P_{A}  ]}{d_S(d_S+1) d_A (d_A + 1)}\\
&=2 \frac{d_S-1}{(d_S+1)(d_A+ 1)}
\end{align}
which, for large dimension $d_S$ returns an average localizable coherence scaling as $\sim 2/d_A$. In addition, if also $d_A \approx d_S$ we obtain a result that is twice as large as for the average localizable coherence by tracing out, Eq.~\eqref{tr_ave_coh}. On the contrary, in situations where $d_A\gg d_S\gg 1$, this measurement protocol yields a much lower localizable coherence on the average factorized state.

At this point, we want to set the stage so that the notion of 
localizable coherence can be used to describe different quantum many-body systems. In the case of a chain of $d_{\loc}$-level systems, the total Hilbert space $\mathcal H $ is the tensor product of local Hilbert spaces corresponding to a single spin system, that is,  $\mathcal H = \mathcal H_{\loc} ^{\otimes n}\simeq  (\mathbb C^{d_{\loc}} )^ {\otimes n}$; and similarly the Hilbert spaces $\mathcal H_S$ and $\mathcal H_A$ can be further decomposed in tensor products of the single spins. Let us consider  $\mathcal H = \mathcal H_{loc} ^{\otimes n}$  for $n = n_S + n_A$, i.e., $n_S$ and $n_A$ correspond to the number of spins in the ``system'' and ``ancillary'' partitions, respectively. We denote $\dim (\mathcal H_{\loc}) = d_{\loc}$.

In such systems, it is interesting to consider states that are factorized in all the spins, or in blocks of spins.
A completely factorized state has the form $U_1 \otimes \dots \otimes U_n (\ket{0} \bra{0})^{\otimes n} U_1^\dagger \otimes \dots \otimes U_n^\dagger $ where each of the $U$'s is Haar i.i.d. and $\ket{\psi_0}$ is any pure state in $\mathcal H_{\loc}$ that we take as reference state. Denoting $\omega \coloneqq \ket{0} \bra{0}^{\otimes n}$ the completely factorized state can be generically expressed as $\omega_{\tilde{U }}= \otimes_i U_i \omega  \otimes_i U_i^\dagger$, with $\tilde{U}=\otimes_{i=1}^n U_i $.


We are interested in knowing the average localizable coherence (without post-selection) in completely factorized states. 
We have
\begin{align*}
\overline{C_B^{(S)}(\omega_{\tilde{U }}) }^{\tilde{U}} 
 = \Tr \left[ S   \left( \mathcal D_{B_{A}}^{\otimes 2} - \mathcal D_{B}^{\otimes 2} \right) \prod_{\alpha=1}^n \left( \frac{I + S_{\alpha \alpha'}}{d_{\loc}(d_{\loc}+1)} \right)  \right] 
\end{align*}
where $S_{\alpha\alpha'}$ denotes the swap operation between spins $\alpha$ and
$\alpha'$ (its corresponding in $\mathcal H'$). Expanding the product, we get
\begin{align}
\overline{C_B^{(S)}(\omega_{\tilde{U }}) }^{\tilde{U}}  = \frac{1}{(d_{\loc}(d_{\loc}+1))^n} \left( T_1 - T_2 \right) 
\end{align}
where we set
\begin{align*} 
T_1 & = \Tr \left[ S  \,  \mathcal D_{B_{A}}^{\otimes 2}  \left[ \prod_{\alpha=1}^n \left( I + S_{\alpha \alpha'} \right) \right]  \right] \,\;, \\
T_2 & = \Tr \left[ S  \,  \mathcal D_{B}^{\otimes 2}  \left[ \prod_{\alpha=1}^n \left( I + S_{\alpha \alpha'} \right) \right]  \right]  \,\;.
\end{align*}
For the calculation of $T_1$ we need to count the swap terms that involve indices $\alpha$ that belong in the ``system'' part.  Given a partition of the spins in $(S,A)$, we define 
\begin{align*}
q_{n_S,n_A}^k (l) \coloneqq {{n_S} \choose {l}} {{n_A} \choose {k-l}} 
\end{align*}
which corresponds to the different ways of choosing $k$ out of $n = n_S + n_A$ $d_{\loc}$-level systems such that exactly $l$ of them are in the ``system'' partition. We have
\begin{align*}
T_1 = \sum_{k=0}^n \sum_{l=0}^k (d_{\loc})^{n+l} q_{n_S,n_A} ^k(l)  \,\;,
\end{align*}
since each of the terms with $l$ swaps in the system part contributes with a factor of $(d_{\loc})^{n+l}$.
The $T_2$ term does not differentiate between the subsystems $S,A$, and a similar calculation gives
\begin{align*}
T_2 = \sum_{k=0}^n {{n} \choose {k}} (d_{\loc})^{n} = (2 d_{\loc})^{n}  \,\;.
\end{align*}
Combining the previous expressions, we finally get
\begin{align} \label{average_local}
\overline{C_B^{(S)}(\omega_{\tilde{U }}) }^{\tilde{U}}  = \frac{1}{(d_{\loc}+1)^n} \left( \sum_{k=0}^n \sum_{l=0}^k q_{n_S,n_A}^k (l) d_{\loc}^{\,l} - 2^n \right)
\end{align}

As a simple crosscheck, one can set $n=n_S = 1$, in which case the result collapses to Eq.~\eqref{average_coherence} for $d = d_{\loc}$. 
As we can see, if as input we have product states then the localizable coherence given by the (non-selective) measurement protocol is a viable way of storing coherence in a subsystem.


\section{Applications to quantum many-body systems}

In this section, we apply some of the ideas and results introduced so far to the description of notable quantum many-body states from the coherence point of view. We are interested in states that can be representative of the ergodic phase (as described by the Eigenstate Thermalization Hypothesis~\cite{deutsch1991quantum,srednicki1994chaos,
rigol2008thermalization} (ETH)), of the MBL phase, and of the topologically ordered phases.  We model the ETH state simply like a Haar-random state in the Hilbert space. These states do indeed obey a volume law for the entanglement, and ergodicity ensures that all the states in (a subspace) of the Hilbert space can be reached with equal probability. In order to describe MBL and topologically ordered states, though, we need a bit more work.

\subsection{Localized states}

\begin{figure}[t]
  \centering
    \includegraphics[width=0.9\columnwidth]{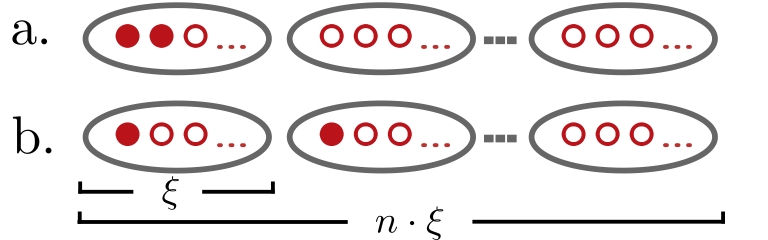}
     \caption{Schematic representation of random states resembling the MBL phase. Each constituent (e.g., spin in a chain) corresponds to a Hilbert space with dimension $d_{\loc}$ and is represented by a red circle. The gray bubbles denote the action of independent randomizing unitaries. The system consists of two constituents (pair of filled red circles), acted upon by either (a) the same unitary or (b) by two different ones.} \label{fig:bubbles}
\end{figure}

Here we want to describe states that can be representative of the MBL phase. Such states should be weakly entangled and feature an area law. However, within the correlation length $\xi$ associated with the localized phase, the states can be highly entangled. We will hence consider as representatives of MBL phase states consisting of products of bubbles of length $\xi$, such that the constituents (e.g., spins) within each bubble are highly entangled but the splitting in-between the different bubbles enforces an area law for the entanglement, see \autoref{fig:bubbles}. We model such states as the tensor product of states that are Haar random within the correlation length $\xi$. These states are thus extremely localized as there is no entanglement at all between one bubble and another.

Equipped with the results from the previous section, we want to perform the average over the localizable coherence on the above described states.  Consider $N = n \cdot \xi$ identical systems that are acted upon by $n$ i.i.d. unitaries, each acting on $\xi$ systems. Each system has a (fixed) dimension $d_{\loc}$, so that $d = d_{\loc}^{n \xi}$. As an example, consider  quantum states $\Phi$ of a spin one-half chain, so that the local Hilbert space at the site $i$ is $\mathbb C^2$ and $d_{\loc}=2$. A localized state with correlation length $\xi$ is a state that resembles a product state of a system with $\xi$ spins, that is, $\ket{\Phi} = \otimes_{k=1}^n \ket{\phi}_k$ with $\ket{\phi}_k\in\mathbb (\mathbb C^{2})^{\otimes\xi}$. In other words, this state is the product of $n$ bubbles of spins, each containing $\xi$ spins.  Within each bubble, the state can be highly correlated and highly entangled.

This setup is convenient to study some interesting class of many-body quantum states like many-body localized quantum states. Indeed, by averaging over bubbles of length $\xi$, we obtain a state that is highly correlated (and entangled) within each bubble, but that is factorized over the bubbles. This state can be used as a reference state for the quantum many-body localized phase. On the other hand, the global Haar state is a representative of the ETH phase, at infinite temperature. Now, imagine to consider the system $S$ made of two parts, so that $N_S = 2$ and $N_A = N - 2$. We ask whether it makes any difference for the localizable coherence whether these two parts are close to each other. Obviously, in the ETH case, it does not, as the global Haar measure does not see any internal structure of the states. However, in the case of averaging over the bubbles, there are two distinct cases, see \autoref{fig:bubbles}.
\begin{enumerate}[a.]
\item The two constituents of the systems are acted upon by the same unitary, i.e., they are within the same bubble. Notice that hence, in this case, it must be $\xi \ge 2$ and $n \ge 1$.

\item The two constituents of the systems are acted upon by two different unitaries, i.e., they belong in two separate bubbles. In this case $\xi \ge 1$ and $n \ge 2$.
\end{enumerate}
As in the previous section, one can write
\begin{multline*}
\overline{C_B^{(S)}\left(  U_1 \otimes \dots \otimes U_n (\ket{0} \bra{0})^{\otimes n \xi}\, U_1^\dagger \otimes \dots \otimes U_n^\dagger  \right)}^{\,U_1,\dots,U_n} \\
 = \frac{1}{(d_{\loc})^{n \xi}(d_{\loc}^\xi+1)^n} \left( T_1 - T_2 \right) 
\end{multline*}
where we set
\begin{align*}
T_1 & = \Tr \left[ S  \,  \mathcal D_{B_{A}}^{\otimes 2}  \left[ \prod_{\alpha=1}^n \left( I + S_{\alpha \alpha'} \right) \right]  \right] \\
T_2 & = \Tr \left[ S  \,  \mathcal D_{B}^{\otimes 2}  \left[ \prod_{\alpha=1}^n \left( I + S_{\alpha \alpha'} \right) \right]  \right] 
\end{align*}
One can now perform a similar calculation, counting the number of terms with different contributions. For the case (a), we have
\begin{multline}
\overline{C_B^{(S)}\left(  U_1 \otimes \dots \otimes U_n (\ket{0} \bra{0})^{\otimes n \xi} \, U_1^\dagger \otimes \dots \otimes U_n^\dagger  \right)}^{\,U_1,\dots,U_n} \\  
=  \left( \frac{2}{d_{\loc}^\xi +1} \right)^n  \frac{d_{\loc}^2 - 1}{2}  \,\;,
\end{multline}
while for case (b)
\begin{multline}
\overline{C_B^{(S)}\left(  U_1 \otimes \dots \otimes U_n (\ket{0} \bra{0})^{\otimes n \xi} \, U_1^\dagger \otimes \dots \otimes U_n^\dagger  \right)}^{\,U_1,\dots,U_n} \\ 
=  \left( \frac{2}{d_{\loc}^\xi +1} \right)^n \frac{d_{\loc}^2 + 2 d_{\loc}  - 3}{4}  \,\;.
\end{multline}

As we can see, the ratio between the localizable coherence in the two cases is $2 \times (d_{\loc}^2 - 1)/(d_{\loc}^2 + 2 d_{\loc}  - 3)$. This number is $6/5$ for qubits, where $d_{\loc}=2$, and converges to $2$ for large local Hilbert space dimension. In other words, there is more localizable coherence if the system is inside the localized bubble (a) than if the system is made of two parts far away (b). In this sense, the localizable coherence captures the fact that the state has the local structure of bubbles. The representative for the ETH state is the random Haar state for which, on the other hand, there is no difference in where and how the system $S$ is located. Of course, this is a cartoon simplified picture of the structure of ETH and MBL states as MBL states are not made exactly of disentangled bubbles (rather, bubbles entangled with area law with each other) and ETH states are not Haar-random but share with Haar-random volume law for entanglement. This result, though,  suggests that localizable coherence could be used as a tool to detect the ETH-MBL transition~\cite{dhara2020quantum}.


\subsection{Toric code} \label{sec:toric}
In this section, we show how the notion of localizable coherence can capture topological features of topologically ordered quantum states like the ground state of the string-net states,  quantum double models, or quantum lattice gauge theories~\cite{kitaev2003fault, hiz, hammawen}. Let us first show how the localizable coherence in the reduced density matrix does have a topological character. In these theories, the reduced density matrix $\rho_S$  of the ground state has a flat spectrum $\{ |c_a|^2 \}_a$~\cite{hammawen}. Denote $\tilde{r}$ the rank of $\rho_S$. In this case, one has $c_{2,B_S}(\rho_S) = \tilde{r}^{-1} -  \tilde{r}^{-2} \sum_{k} \left( \sum_a   |\langle \xi_a| k\rangle |^2 \right)^2$. Choosing a mutually unbiased basis one obtains $c_{2,B_S}(\rho_S) = \tilde{r}^{-1}-d_S^{-1} $. Similarly, for the same states with flat entanglement spectrum,  the $l_1$--norm of coherence reads $c_{1,B_S} \left( \rho_S \right) = \tilde{r}^{-1} \sum_{k\ne k'} \Big| \sum_{a=1}^{\tilde{r}  }\langle\xi_a |k\rangle\langle k'|\xi_a\rangle  \Big|$.

Notice that, in the particular case of the toric code (or quantum lattice gauge theories), the rank $R$ is not full; first of all because there is area law, and then because  there are prohibited configurations on the boundary of the system. In fact, that correction is the topological ``missing'' entropy $\log \gamma$~\cite{hammawen} and one has $\tilde{r}= d_{\partial S}/\gamma$. If one chooses as subsystem $S$ a thin region (without bulk)~\cite{halasz}, then $\partial S = S$ and we obtain  $c_{2,B_S} \left( \rho_S \right) = (\gamma-1)/d_{\partial S}$. This has to be compared with other states with flat entanglement spectrum that are not topologically ordered, where $\gamma = 1$, hence the coherence vanishes. In this sense, the previous formula shows a topological coherence.

As a second application, we show that the localizable coherence by measurement can reveal topological properties. For this purpose let us focus on the toric code~\cite{toriccode}. In order to understand the measurement protocol, we need to go into the details of the model. The toric code with spins one half on the bonds of a $N\times N$ square lattice with periodic boundary conditions is described by the model Hamiltonian
\begin{align}
H = -U \sum_n A_n - J \sum_p B_p  \,\;.
\end{align} 
where the `star' operator $A_n= \prod_{l\in n} \sigma^x_l$ flips all the spins (in the $z-$ basis) extruding from a vertex $n$ and the `plaquette' operator $B_p=\prod_{l\in p} \sigma^z_l$ operates with $\sigma^z$ on all the spins around a plaquette $p$. 
Denoting by $G$ the group generated by the star operators $A_n$, that is, the set obtained by all the possible products of operators $A_n$, 
the ground space of the toric code Hamiltonian~\cite{hiz, hiz3} can be written as the span of the vectors
\begin{align}
\ket{\psi_0} = \sum_{i,j \in \{ 0,1 \}} \alpha_{ij} (W_1^x )^i (W_2^x )^j \left| G \right|^{-1/2} \sum_{g \in G} \ket{g}  \,\; ,
\end{align}
where $W_1^x,W_2^x$ correspond to the product $\sigma_i^x$ operators over horizontal and vertical non-contractible loops of the torus. It follows that the ground space has degeneracy $4$. Considering the quantum coherence of the above ground states with respect to the product $ \sigma_i^z$ eigenbasis, one identifies two contributions to it: (i) from the coherent superposition of the 4-fold degenerate ground states (i.e., due to the $\alpha_{ij}$ coefficients), and (ii) from the equal superposition of terms in the group $G$.

We will now show that the 2-norm localizable coherence by measurement $C_{\ave,B} ^{(S)}$ can differentiate between the aforementioned two types of coherence. Moreover, the topology of the region where measurements are performed will play a role in the result, revealing a topological character.

\begin{figure}[t]
    \centering
        \includegraphics[width=0.4\textwidth]{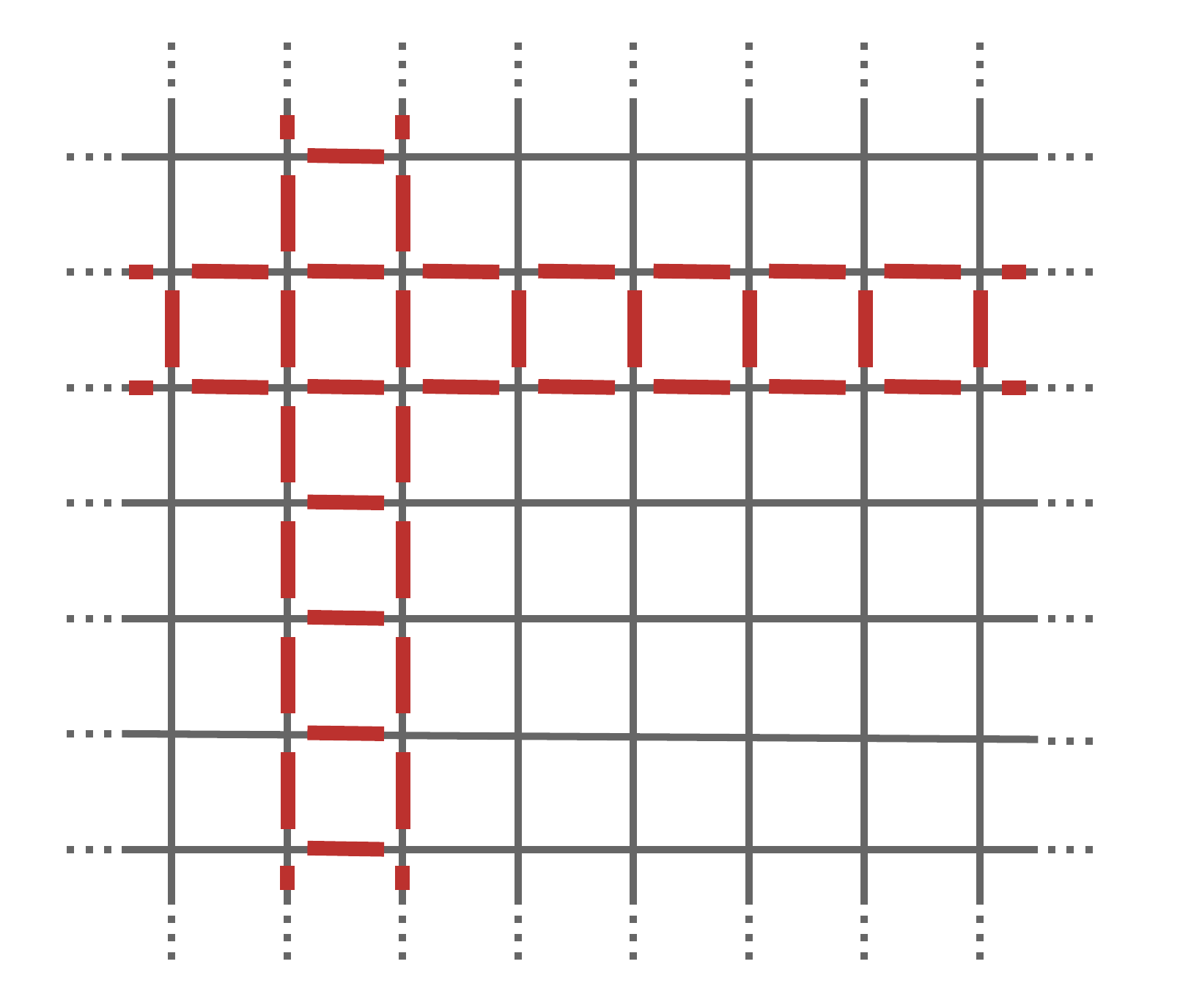} 
        \caption{Toric code; spins reside on the edges. The system (red) non-contractible region consists of two strips that wrap around the torus and meet at a single plaquette. In the ancilla (gray) region spins are measured over the local $\sigma^z$ basis, and the region is contractible.}
\label{tc_1}
\end{figure}

For this purpose, it is instructive to first analyze the scenario where one performs orthogonal and selective $\sigma_i^z$ measurements for all spins except those belonging to two strips of plaquettes, one horizontal and one vertical (see \autoref{tc_1}). We will refer to the ancillary part consisting of the measured spins as $A$, while we consider the complement to be the system $S$. In fact, the exact shape of the two regions is not going to matter for the considerations that follow, except from the fact that the region $S$ is topologically non-contractible.

After the measurement, the resulting state is of the form
\begin{align}
\ket{\psi_0} \mapsto \ket{\psi'_0} \propto  \left( I_S\otimes\ket{h_A}\bra{h_A} \right) \ket{\psi_0} \,\;.
\end{align}
For a given measurement result $h_A$ in region $A$, let
\begin{align}
\ket{h} = \ket{h_S}\ket{h_A} \,, \quad h \in G
\end{align}
be a ``completion''.\footnote{Notice that, although the part $\ket{h_S}$ might not be unique, a completion always exists since we have assumed that the measurement result $\ket{h_A}$ occurred.} Given that the region $S$ is non-contractible, and the group average is invariant under group multiplication, one can write
\begin{align*}
\ket{\psi_0} &\propto \sum_{i,j} \alpha_{ij} (W_1^x )^i (W_2^x )^j  \sum_{g \in G} \ket{g}\\
 &= \sum_{i,j} \alpha_{ij}   \sum_{g \in G} \left[ (W_1^x )^i (W_2^x )^j \ket{g_S} \right] \ket{g_A}  \\
 &= \sum_{i,j} \alpha_{ij}   \sum_{g \in G} \left[ (W_1^x )^i (W_2^x )^j h_S g_S\ket{0_S}  \right] h_A g_A\ket{0_A} 
\end{align*}
and hence also
\begin{align*}
\ket{\psi_0'} \!\propto \! \sum_{i,j} \alpha_{ij}  \! \sum_{g \in G} \left[ (W_1^x )^i (W_2^x )^j h_S g_S\ket{0_S}  \right] \ket{h_A} \!\bra{0_A}  g_A  \ket{0_A} .
\end{align*}
The only surviving term is for $g \in G$ such that $g_A = id_A$. One can formally define the subgroup $G_S \coloneqq \{ g \in  G : g= g_S \otimes id_A \}$. Therefore after the measurement the normalized state is
\begin{align}
\ket{\psi_0'} = \left| G_S \right|^{-1/2} \sum_{i,j} \alpha_{ij}   \sum_{g \in G_S}  (W_1^x )^i (W_2^x )^j \ket{h_S g_S} \ket{h_A}  \,\;. \label{tor_post_meas_state}
\end{align} 
It is now important to observe that $ G_S $ depends on the geometry of the partitioning $(S,A)$ but not on the particular measurement result $h_A$. For instance, for the region as in \autoref{tc_1}, $\left| G_S \right| = 2^4$.

We are now ready to calculate $C_{\ave,B} ^{(S)} (\ket{\psi_0}\!\bra{\psi_0} )$ for $B$ the product $\sigma_i^z$ eigenbasis. 
A straightforward calculation for the coherence of the post-selected state gives
\begin{align}
c_{2,B_S} \left( \Tr_A \ket{\psi_0'}\! \bra{\psi_0'} \right) = 1 - \frac{1}{\left| G_S \right|} \sum_{i,j} \left| \alpha_{ij} \right|^4 
\end{align}
which is also independent of $h_A$. Therefore, under the sole assumption that $S$ is non-contractible, one obtains
\begin{align} \label{torus_non_contr}
C_{\ave,B} ^{(S)} (\ket{\psi_0 }\!\bra{\psi_0}) = 1 - \frac{1}{\left| G_S \right|} \sum_{i,j} \left| \alpha_{ij} \right|^4  \,\;.
\end{align}

The resulting coherence $C_{\ave,B} ^{(S)} (\ket{\psi_0 }\!\bra{\psi_0})$ is sensitive to the superposition within the 4-dimensional ground state subspace, while the contribution of the corresponding group $G$ is through the factor $\left| G_S \right|$ that depends on the geometry. More importantly, it reveals topological features of the $(S,A)$ partitioning. To see this, let us consider what happens in the opposite case where $S$ is contractible. Then, a measurement on $A$ always collapses the 4-fold superposition due to the $\alpha_{ij}$, i.e., from the measurement result on $A$ one can infer the definite values of $i,j \in \{ 0,1 \}$; this is done just by analyzing whether or not the obtained configuration contains non-contractible loops along the horizontal and vertical directions. For a measurement result $(W_1^x )^i (W_2^x )^j\ket{h_A}$, the post-measurement state now becomes
\begin{align}
\ket{\psi_0''} = \left| G_S \right|^{-1/2} \sum_{g \in G_S}   \ket{h_S g_S} (W_1^x )^i (W_2^x )^j\ket{h_A}  \,\;, \label{tor_post_meas_state_2}
\end{align}
where the values of $i,j$ are fixed depending on the measurement outcome. Once again, the coherence of the resulting state $c_{2,B_S} \left( \Tr_A \ket{\psi_0''}\! \bra{\psi_0''} \right)$ is independent of $h_A$ and also $i,j$, therefore one obtains for contractible $S$,
\begin{align}\label{torus_contr}
C_{\ave,B} ^{(S)} (\ket{\psi_0 }\!\bra{\psi_0}) = 1 - \frac{1}{\left| G_S \right|} \,\;.
\end{align}

The analogous expressions when $S$ is only contractible along only one of the horizontal/vertical directions can be obtained similarly. Eqs.~\eqref{torus_non_contr} and \eqref{torus_contr} therefore show that the localizable coherence by measurement $C_{\ave,B} ^{(S)} (\ket{\psi_0 }\!\bra{\psi_0})$ is sensitive to the superposition over the 4-dimensional toric code groundspace, but only if the $S$ region is non-contractible along the corresponding direction.


\section{Conclusions and Outlook} 
In this paper, we have addressed the question of quantifying coherence in a composite quantum system where a notion of locality is imposed by a tensor product structure. We have put forward a notion of localizable coherence as the coherence that is obtainable in a subsystem of a composite quantum system after either disregarding or by measurement on the rest of the system, that serves as an ancilla. We have computed the average localizable coherence  over the Hilbert space, including over different factorizable states. It results that measurement aided localizable coherence is more efficient than simply tracing out the ancillary system, as this would result in strong decoherence. As an application, we have shown that localizable coherence can distinguish between topological characters of many-body quantum states, for example, the toric code.

One of the examples discussed suggests that localizable coherence could be a useful quantity to characterize the  ETH-MBL transition. This connection is explored in more detain in~\cite{dhara2020quantum}.

In perspective, localizable coherence can potentially provide useful insights in situations where one wants to understand the role of coherence in quantum systems with a tensor product structure, for instance, in quantum thermodynamics of composed systems, like quantum batteries, or in the role played by coherence in operator spreading~\cite{spreading}, scrambling, and the transition to quantum chaotic behavior~\cite{quantumchaos} signaled by out-of-time order correlation functions~\cite{complexitybydesign}. Further investigation of these subjects, possibly under the lens of localizable coherence, provides directions for future research.

\acknowledgments

P.Z. acknowledges partial support from the NSF award PHY-1819189.  Research was funded by the Deutsche Forschungsgemeinschaft (DFG, German Research Foundation) under Germany's Excellence Strategy -- EXC-2111 -- 390814868. This research was (partially) sponsored by the Army Research Office and was accomplished under Grant Number W911NF-20-1-0075. The views and conclusions contained in this document are those of the authors and should not be interpreted as representing the official policies, either expressed or implied, of the Army Research Office or the U.S. Government. The U.S. Government is authorized to reproduce and distribute reprints for Government purposes notwithstanding any copyright notation herein.

\bibliography{refs}

\appendix

\section{Optimal coherence basis for $C_{B} ^{(S)}(\rho)$ and $C_{\ave,B} ^{(S)} (\rho)$} \label{appendix}

Here we determine the coherence basis $B_S$ such that, for fixed $B_A$ and $\rho$, each of the quantities
$C_{B} ^{(S)}(\rho)$ and $C_{\ave,B} ^{(S)} (\rho)$ become maximal for the coherence quantifier $c_{2,B}$, under the assumption that $\{ \rho'_{S,i} \}_i$ are mutually commuting. Let $B_S'$ be a basis that simultaneously diagonalizes $\{ \rho'_{S,i} \}_i$. We will show that the optimal choice is, in both cases, a basis $B_S$ that is unbiased to $B_S'$.

Let us begin with $C_{B} ^{(S)}$. We have
\begin{align*}
C_{B} ^{(S)}(\rho) = \Pur\left[\mathcal D_{B_A} (\rho) \right] - \Pur\left[ \mathcal D_{B_S}  \mathcal D_{B_A} (\rho) \right]
\end{align*}
and hence we are looking for the choice of $B_S$ that minimizes the second term. By the mutually commuting assumption and setting $B_S' = \{ \nu_i \}_i$,
\begin{align*}
\sigma \coloneqq \mathcal D_{B_A} (\rho) = \sum_{j} p_j \rho'_{S,j} \otimes \omega_j = \sum_{ij} q_{ij} \nu_i \otimes \omega_j \,\;,
\end{align*}
where $\{ q_{ij} \}_{ij}$ are elements of a (bipartite) probability distribution. We can hence write
\begin{align*}
\Pur\left[ \mathcal D_{B_S}  (\sigma) \right] & = \Pur\left[ ( \mathcal U^\dagger \otimes \mathcal I)  \mathcal D_{B_{S}'} ( \mathcal U \otimes \mathcal I ) (\sigma) \right] \\ & = \Pur\left[\mathcal D_{B_{S}'} ( \mathcal U \otimes \mathcal I ) (\sigma) \right] 
\end{align*} 
where $\mathcal U$ is a unitary that connects the bases $B_{S}'$ and $B_S$, and the last step follows since purity depends only on the spectrum. In other words, optimizing the basis $B_S$ is equivalent to fixing the basis to $B_{S}'$ and optimizing the unitary $\mathcal U$.

The above step reduces the problem to a classical one, since by evaluating the above expression one gets
\begin{align} \label{app}
\Pur\left[ \mathcal D_{B_S}  (\sigma) \right] = \Pur\left[ (M \otimes I) q  \right] \,\;,
\end{align}
where $M^{(U)}$ is the unistochastic matrix~\cite{bengtsson2017geometry} with elements $M^{(U)}_{ki} = \Tr \left[ \nu_k \mathcal U (\nu_i) \right]$, while the purity on the RHS is that of a probability vector (and not of a density matrix). In other words, now the problem reduces to specifying the unistochastic matrix $M$ that minimizes the purity of a fixed probability vector $q$ as in Eq.~\eqref{app}.

The answer to the above is easily obtained using the theory of majorization~\cite{bhatia2013matrix}. It amounts to recalling that purity is a Schur-convex function and hence the action of a bistochastic matrix monotonically decreases the purity. The minimum is therefore obtained for $M^{(U)}_{ki} = 1/d_S$ for all vectors $q$, which corresponds to $B_S$ and $B_{S}'$ being unbiased. Notice that the choice is independent of $q$, which is a consequence of the simplifying assumption about mutual commutativity.

Regarding $C_{\ave,B} ^{(S)} (\rho)$, first notice that the probability $\{p_i\}$ in Eq.~\eqref{average_measure} is independent of $B_S$. In addition, each of the $c_{B_S} (\rho'_{S,i})$ obtain their maxima simultaneously also for $B_S$ and $B_{S}'$ being unbiased. This also follows from the above arguments by setting $d_A = 1$.


\end{document}